\documentclass{article}
\usepackage{graphicx}
\usepackage{latexsym}

\begin{document}
\begin{flushright}
UM-P 031-2000\\
RCHEP 007-2000
\end{flushright}
\begin{center}
{\huge Signature of Randall-Sundrum Quantum Gravity model in
$\gamma\gamma$ scattering in the TeV range.}\\
\hspace{10pt}\\
S.R.Choudhury\footnote{src@ducos.ernet.in} \\
{\em Department of Physics, Delhi University, Delhi, India},\\
A. Cornell\footnote{a.cornell@tauon.ph.unimelb.edu.au}
and G.C. Joshi\footnote{joshi@physics.unimelb.edu.au}\\
{\em School of Physics, University of Melbourne,}\\
{\em Parkville, Victoria 3108, Australia}\\
\hspace{10pt}\\
$22^{nd}$ of July, 2000
\end{center}
\hspace{10pt}\\   
\begin{abstract}
We examine the implications of the Randall-Sundrum gravity models on
$\gamma\gamma$ scattering in the TeV range.
\end{abstract}  

\indent It has recently been proposed that the scale of weak
interactions $M\sim 1TeV$ is the only fundamental scale of energy in
the description of particle interactions.  The conventional scale of
gravitational interactions, the Planck scale $M_{pl} \gg M$ is no
longer fundamental but arises out of a Kaluza-Klein compactification
of a higher dimensional space-time into four dimensions.  In the first
such proposals of weak scale quantum gravity (WSQG), one starts with
$(4+n)$-dimensional space-time with the standard model (SM) fields
confined to a three-brane that we live in \cite{one}.  Gravity, on the
other hand, is postulated to propagate in the full space-time
including the $n$ extra dimensions which are assumed to be compact of
size $r_c$.  For distances $r\ll r_c$, gravity is no different from
other SM interactions, but for $r\gg r_c$ the net effect of
propagation of gravitons in the extra compact dimensions is to make
strength of the gravitational interactions from $1/M$, via the
relation:
\begin{eqnarray}
M_{pl}^2 \sim M^{n+2}\cdot r_c^2 .
\end{eqnarray}
For $M\sim 1 - 10TeV$, this last equation gives as low as $1mm$, which
is the current experimental limit of validity of Newtonian gravity
law.  This scenario has interesting phenomenological implications in
TeV-scale physics \cite{two} as well as in the precise values of
certain experimentally measurable low energy parameters like $(g-2)$
of the muon \cite{three}.  The existence of the $n$-compact dimensions
of course implies that the graviton is accompanied by a tower of
massive cousins.  These will be individually gravitationally coupled
but added up, these effectively give rise to interactions comparable
or even greater than in strength of weak interactions at sufficiently
high energies in the TeV scale.  Amongst the TeV-scale processes,
clearly the ones that will be most sensitive to these `new'
interactions are the ones which are sufficiently suppressed in the SM.
The scattering process $\gamma\gamma\to\gamma\gamma$ is one such
example since in the SM, the amplitude for this process is of
$O(\alpha^2)$\cite{four}.  This is also a process is which there is a
possibility of obtaining TeV-scale data in the near future and thus is
worth investigating.  Such investigations in the context of the
ADD-model just outlined have been done in \cite{five} and
\cite{six}.\\
\indent An alternative scenario of compactification has recently been
proposed by Randall and Sundrum \cite{seven} where, the hierarchy of
scales $M_{pl}$ and $M$ are generated in a different way.  The
space-time is now a 5D-non factorizable geometry, with a compact
angular co-ordinate $\phi$ ranging from $|\phi|=0$ to $\pi$.  Two
3-branes with opposite tensions reside at the points $\phi=0$ and
$\phi=\pi$ and the resultant 4-d metric has the form
\begin{eqnarray} 
ds^2 = e^{-2\sigma(\phi)}\eta_{\mu\nu}\cdot dx^{\mu} dx^{\nu}+ r_c^2
d\phi^2,
\end{eqnarray}
with $\mu,\nu=0,1,2,3$ and $\sigma(\phi)=kr_c|\phi|$, $r_c$ being the
compactification scale.  Our universe in this scenario is at
$\phi=\pi$ and because of the exponential `warp' factor in the
metric, physical masses in our 4-d world are related to the
fundamental scale $M_0$ of the theory by the relation
$M_{phys}=exp(-kr_c\pi) M_0$.  This then is the reason for the weak scale
arising with a value of $kr_c\sim 10$.  Further, because of this
compact dimension, gravitons will occur as a tower of particles.  The
zero mode is uniform in $\phi$ and behaves like an ordinary graviton
with a 4-d $M_{pl}$ scale; the excitations are massive and couple with
matter with a scale $\Lambda=exp(-kr_c)M_{pl}$ which is of the order
of $1TeV$.  The net Lagrangian of the coupling of this tower of
gravitons with matter can then be expressed as a Lagrangian
\cite{eight}:
\begin{eqnarray}
L = -\left(\frac{1}{M_{pl}}\right)T^{\alpha\beta}(x)
h^{(0)}_{\alpha\beta}(x) - \left(\frac{1}{\Lambda}\right)
T^{\alpha\beta}(x) \sum_{n=1}^{\infty}h^{(n)}_{\alpha\beta}(x).
\end{eqnarray}
The scenario in the Randall-Sundrum (RS) version of weak scale quantum
gravity is thus substantially different from the ADD one, where the
excited gravitons effectively form a continuum.  In contrast the
RS-excitations form a discrete set with well defined and calculable
spectrum.  We of course have no hint where exactly will the
RS-excitations lie in the TeV-scale.  However, choosing the mass of
the lowest excitation as a parameter, it is possible to calculate the
spectrum and this is what was done in the first paper on the
phenomenology of the RS-model \cite{nine}.  We will also consider the
mass of the first RS-resonance as a parameter and investigate the
behaviour of $\gamma\gamma$-scattering amplitude in this scenario.
This is thus an extension of the investigations in references by
Davoudiasl and by us, where changes in $\gamma\gamma$ cross sections
arising from weak scale quantum gravity were investigated in the ADD
scenario \cite{five}\cite{six}.  Should there be indications of
departure from the SM in experiments in the near future, the
difference between the two available pictures would be useful.  This
present note addresses thus the question of RS-scenario implication
for the $\gamma\gamma$ scattering.\\
\indent We shall assume, that as in the first paper on
RS-phenomenology, that the first RS-resonance occurs at a value at 600
GeV; changes in that would only shift the graphs but not the
behaviour.  Further, arguments based on string theory tells us that
the parameter $k/M_{pl} \sim 10^{-2}$.  We will therefore investigate
the phenomenology in a range of $k/M_{pl}$ around that.\\
\indent We record first the standard model (SM) amplitudes for
$\gamma\gamma$ scattering.  This is dominated by the W-loop and we
take a region of phase space such that $s$, $|t|$ and $|u|$ are much
greater than $M_w^2$ (that would limit the scattering angle in the
c.m. to be not too close to either $0$ or $\pi$, which of course is
experimentally very convenient).  Under these approximations, the SM
expressions for the helicity amplitudes are, in terms of the standard
Mandelstam variables $s$, $t$, $u$:
\begin{eqnarray}
M_{++++}(s,t,u) = \left(-i\cdot 16\pi\alpha^2\right) \cdot \left( \log
\left|\frac{u}{M_w^2}\right|\cdot\frac{s}{u} + \log 
\left|\frac{t}{M_w^2}\right|\cdot\frac{s}{t}\right)
\end{eqnarray}
\begin{eqnarray}
M_{+-+-}(s,t,u) & = & \left(-i\cdot 12\pi\alpha^2\frac{s-t}{u}\right)
+ \left(i\cdot \frac{8\pi\alpha^2}{u^2}\right) + \left(4u^2 -
3s\right) \left(\log\left|\frac{t}{u}\right|\right) \nonumber \\
& & - \left(i\cdot 16\pi\alpha^2\right) \left(\frac{u}{s} \log
\left|\frac{u}{M_w^2}\right| + \frac{u^2}{st}\log
\left|\frac{t}{M_w^2}\right|\right) 
\end{eqnarray}
\begin{eqnarray}
M_{+--+}(s,t,u) = M_{+-+-}(s,t,u)
\end{eqnarray}
\indent All other amplitudes, not related to these are negligible in
the region considered.\\
\indent The SM-amplitudes have to be added to the contributions
arising out of the exchange of RS-resonances in the $s$-, $t$- and
$u$-channels.  If we assume that the first of the resonances is in the
sub-TeV range (assumed $\sim 600GeV$ in our calculation), then up to
about $\sqrt{s} = 1TeV$, it is a good approximation to consider the
exchange of the lowest RS-resonance only and we will do that.  Using
the interaction given in equation (3), the RS-resonance contributions
to the amplitudes work out to be:
\begin{eqnarray}
M_{++++}^{RS} & = & \left(-\frac{s^2}{\Lambda^2}\right) \left[D(t) +
D(u) \right] \\
M_{+-+-}^{RS} & = & \left(-\frac{u^2}{\Lambda^2}\right) \left[D(s) +
D(t) \right] \\
M_{+--+}^{RS} & = & \left(-\frac{t^2}{\Lambda^2}\right) \left[D(s) +
D(u) \right] 
\end{eqnarray}
where $D(t)= (t-m_1^2)^{-1}$, $D(u)=(u-m_1^2)^{-1}$\\
and $D(s)=(s-m_1^2+im_1\cdot\Gamma + \frac{1}{4}\Gamma^2)^{-1}$.\\
$m_1$ is the mass of the first RS-resonance and $\Gamma$ its width,
which can be calculated as:
\begin{eqnarray}
\Gamma = \left(\frac{3}{10\pi}\right)\left(m_1\cdot x_1^2\right)
\left(\frac{k}{M_{pl}}\right)^2
\end{eqnarray}
$x_1$ above is the first zero of the Bessel function $J_1(x)$.\\
\indent The total amplitude $M^{SM}+M^{RS}$ is therefore determined in
terms of the parameter $(k/M_{pl})$ for a given value of $m_1$.\\
\begin{figure}
\includegraphics[angle=270,width=12cm]{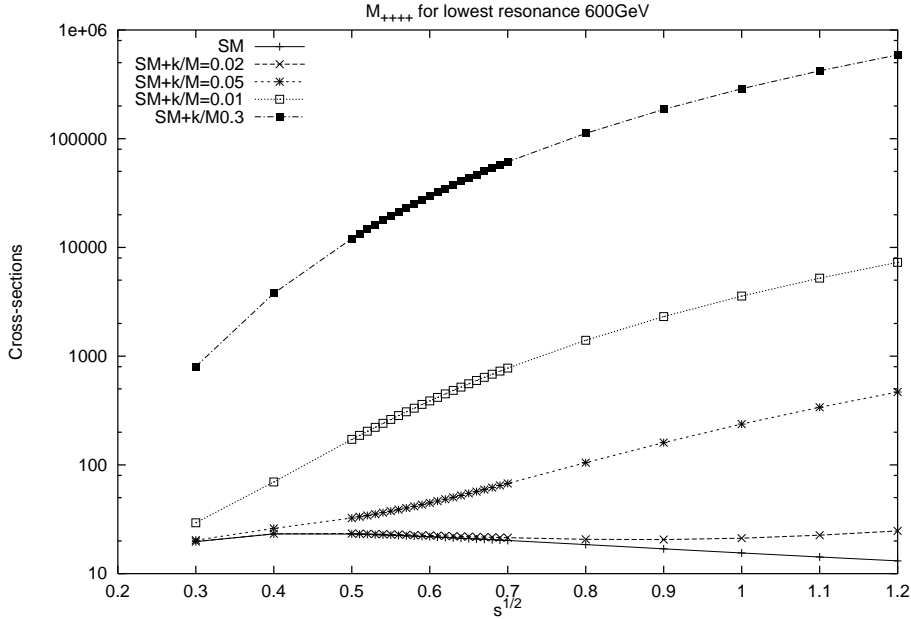}
\caption{Magnitudes of the SM cross-sections and SM plus
Randall-Sundrum contributions for initial helicities ++ for a lowest
resonance of 600GeV.}
\label{fig1}  
\end{figure}
\begin{figure}
\includegraphics[angle=270,width=12cm]{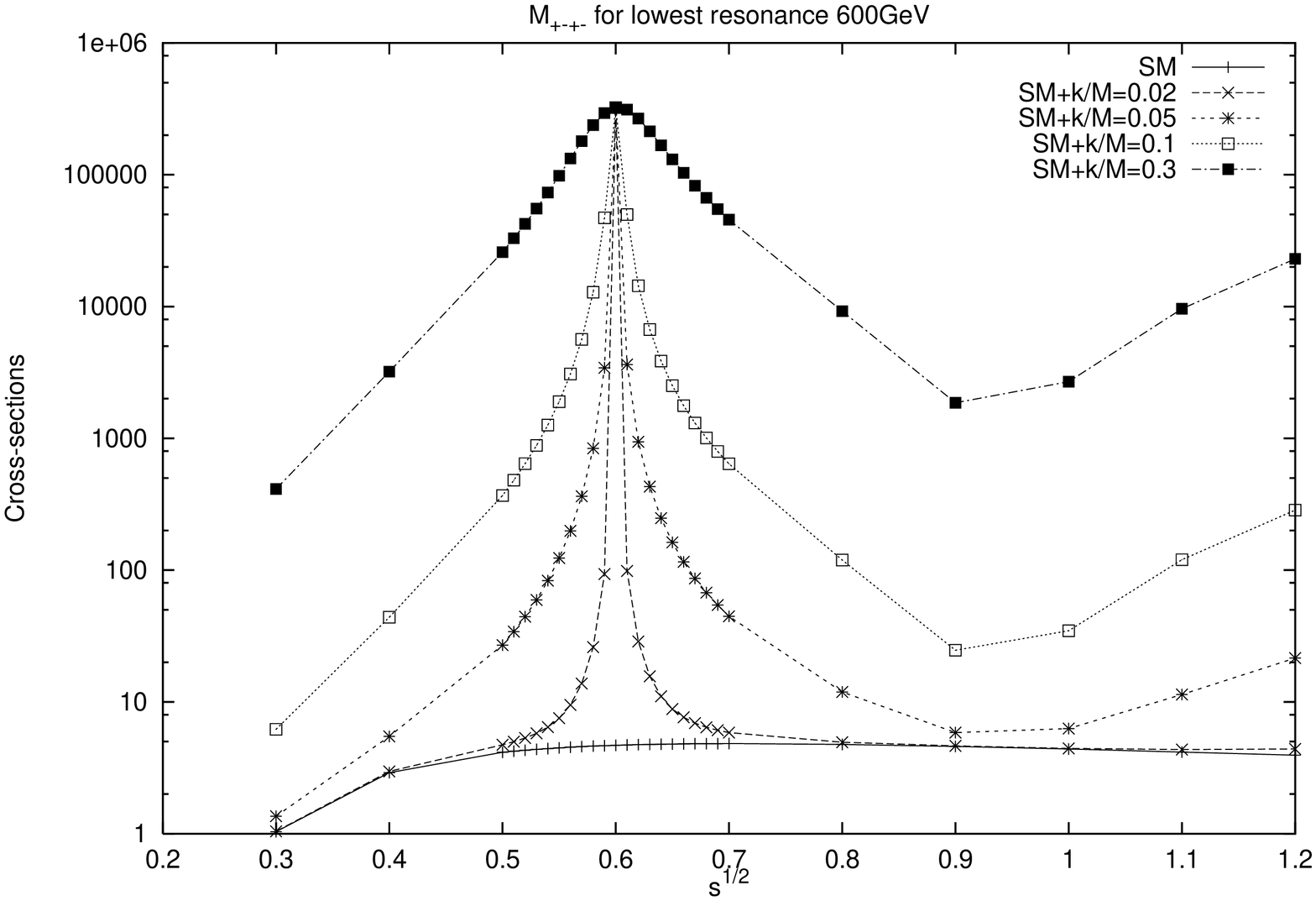}
\caption{Magnitudes of the SM cross-sections and SM plus
Randall-Sundrum contributions for initial helicities +- for a lowest
resonance of 600GeV.}
\label{fig2}  
\end{figure}
\begin{figure}
\includegraphics[angle=270,width=12cm]{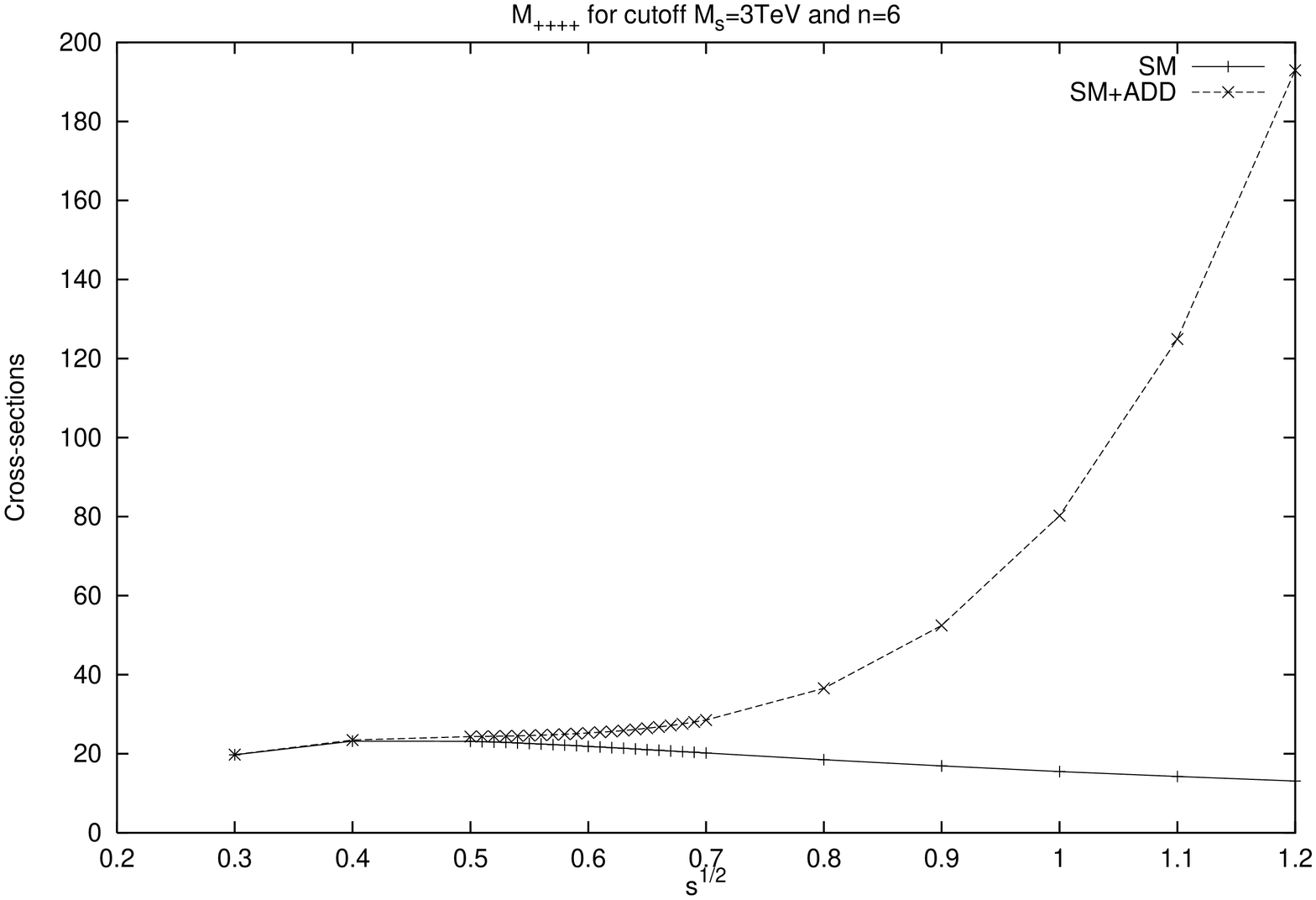}
\caption{Magnitudes of the SM cross-sections and SM plus ADD
cross-sections for initial helicities ++ for a cutoff of $M_s=3TeV$
and n=6.}
\label{fig3}  
\end{figure}
\begin{figure}
\includegraphics[angle=270,width=12cm]{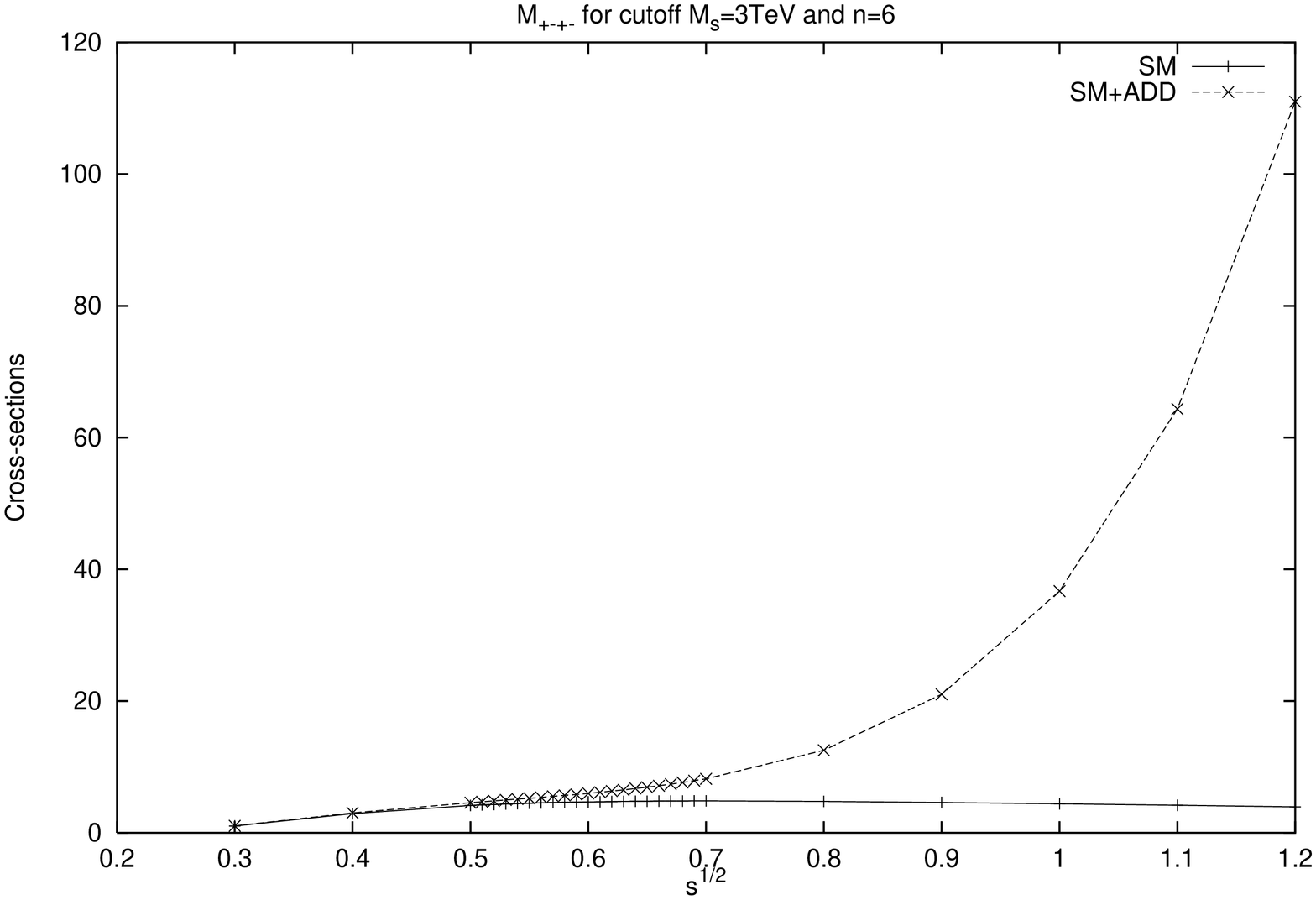}
\caption{Magnitudes of the SM cross-sections and SM plus ADD
cross-sections for initial helicities +- for a cutoff of $M_s=3TeV$
and n=6.}
\label{fig4}  
\end{figure}
\indent Figures 1,2 shows the typical way in which the $\gamma\gamma$
cross-sections for parallel and antiparallel initial helicity states
behave as a function of the c.m. energy.  Also shown are the pure
SM-predictions for comparison.  To see the nature of the difference
between the RS- and ADD-scenario, we have also presented in figures 3
and 4, similar cross-sections for the ADD-scenario with a cut-off
value of $M_s=3TeV$.  We have chosen for the calculations values of
$\sqrt{s}$ up to $1TeV$, hopefully because such energies would be
accessible at NLC for $\gamma\gamma$ processes using scattering of
laser photons from $e^+e^-$ beams.\\
\indent The most striking feature of the RS-scenario, is the huge
enhancement of the cross-sections even for low values of the parameter
$k/M_{pl}$ that we have considered.  In comparison for the pure
fermionic process $e^+e^-\to\mu^+\mu^-$ considered by Davoudiasl, 
Hewett and Rizzo \cite{nine}, the cross section rise is not so
spectacular even for a higher range of values of the parameter
$k/M_{pl}$.  A second feature worth mentioning is the presence of
resonance peaks only in the helicity state $+-$ which should be a very
clear signature should experiments be possible with polarized photons.
The energy dependence of the cross-sections, even in the limited range
considered is also very distinctive of the RS-scenario.  The SM
cross-sections decrease with energy whereas the total (SM+RS)
cross-sections increase mildly with energy.  Of course, these features
would appropriately scaled as the mass of the first resonance takes on
different values, but the qualitative nature of these features will
continue.  It should also be noted that these features are quite
different from the cross-sections obtained by considering the ADD
version WSQG as can be seen by comparison of figure 3 and 4 with their
counterparts, figures 1 and 2.\\
\indent In conclusion, experimental results on $\gamma\gamma$
scattering in TeV scale would not only provide clear indications of
possible departure from SM results not merely as a correction factor
but by a large change in the value of cross-sections.  Not only that,
the values determined and their behaviour as a function of energy will
enable us to distinguish between the two currently available scenarios
of weak scale quantum gravity.

\end{document}